\documentclass[aps, prd, amsmath, floats, floatfix, twocolumn,
superscriptaddress, nofootinbib, showpacs]{revtex4-1}

\usepackage{graphicx}
\usepackage{color}
\usepackage{soul}
\usepackage{url}
\usepackage{bm}         
\usepackage{times}
\usepackage{dcolumn}
\usepackage{bm}
\usepackage{epsf}
\usepackage{amssymb}

\newcommand{\beq}{\begin{equation}}
\newcommand{\eeq}{\end{equation}}
\newcommand{\beqn}{\begin{eqnarray}}
\newcommand{\eeqn}{\end{eqnarray}}

\usepackage{color}

\newcommand{\Caltech}{\affiliation{TAPIR, Walter Burke Institute for Theoretical Physics, MC 350-17,
    California Institute of Technology, Pasadena, California 91125, USA}}

\newcommand{\Cornell}{\affiliation{Cornell Center for Astrophysics and
    Planetary Science, Cornell University, Ithaca, New York, 14853, USA}}
\newcommand{\WSU}{\affiliation{Department of Physics \& Astronomy,
	Washington State University, Pullman, Washington 99164, USA}}

\newcommand{\AEI}{\affiliation{Max Planck Institute for Gravitational Physics (Albert Einstein Institute), D-14476 Potsdam-Golm, Germany}}

\newcommand{\UNH}{\affiliation {Department of Physics, University of New Hampshire, 9 Library Way, Durham NH 03824, USA}}

\newcommand{\GRAPPA}{\affiliation{GRAPPA, Anton Pannekoek Institute for Astronomy and Institute of High-Energy Physics, University of Amsterdam, Science Park 904, 1098 XH Amsterdam, The Netherlands}}

\newcommand{\Nikhef}{\affiliation{Nikhef, Science Park 105, 1098 XG Amsterdam, The Netherlands}}

\begin{document}

\title{Numerical simulations of neutron star-black hole binaries in the near-equal-mass regime}

\author{F. Foucart}\UNH
\author{M.D. Duez}\WSU
\author{L.E. Kidder}\Cornell
\author{S.M. Nissanke}\GRAPPA \Nikhef
\author{H.P. Pfeiffer}\AEI
\author{M.A. Scheel}\Caltech

\begin{abstract}
Simulations of neutron star-black hole (NSBH) binaries generally consider black holes with masses in the range $(5-10)M_\odot$, where we expect to find most stellar mass black holes. The existence of lower mass black holes, however, cannot be theoretically ruled out. Low-mass black holes in binary systems with a neutron star companion could mimic neutron star-neutron (NSNS) binaries, as they power similar gravitational wave (GW) and electromagnetic (EM) signals. To understand the differences and similarities between NSNS mergers and low-mass NSBH mergers, numerical simulations are required. Here, we perform a set of simulations of low-mass NSBH mergers, including systems compatible with GW170817. Our simulations use a composition and temperature dependent equation of state (DD2) and approximate neutrino transport, but no magnetic fields. We find that low-mass NSBH mergers produce remnant disks significantly less massive than previously expected, and consistent with the post-merger outflow mass inferred from GW170817 for moderately asymmetric mass ratio. The dynamical ejecta produced by systems compatible with GW170817 is negligible except if the mass ratio and black hole spin are at the edge of the allowed parameter space. That dynamical ejecta is cold, neutron-rich, and surprisingly slow for ejecta produced during the tidal disruption of a neutron star : $v\sim (0.1-0.15)c$. We also find that the final mass of the remnant black hole is consistent with existing analytical predictions, while the final spin of that black hole is noticeably larger than expected -- up to $\chi_{\rm BH}=0.84$ for our equal mass case.
\end{abstract}

\pacs{04.25.dg, 04.40.Dg, 26.30.Hj, 98.70.-f}

\maketitle

\section{Introduction}
\label{sec:intro}

The first detections of black hole-black hole mergers~\cite{LIGOVirgo2016a} and of one likely neutron star-neutron star (NSNS) merger~\cite{TheLIGOScientific:2017qsa} have shown that gravitational wave (GW) astronomy is now a reality. Electromagnetic (EM) observations~\cite{GBM:2017lvd,2017ApJ...848L..13A} from that NSNS merger, GW170817, have also allowed us to connect NSNS mergers with short-hard gamma-ray bursts, kilonovae, and the production of at least some of the r-process elements.

The determination that GW170817 is most likely a NSNS merger relies on the very reasonable expectation that compact objects in binary systems emitting detectable gravitational wave signals and with masses $M\lesssim 2M_\odot$ are neutron stars. While the presence of at least one neutron star is required by the observation of bright post-merger EM signals, the interpretation of the second object as a neutron star is mostly due to its measured mass. Existing mass measurements for stellar mass black holes in the Milky Way favor black holes masses mostly in the $M_{\rm BH}\sim (5-10)M_\odot$ range (e.g.~\cite{Ozel:2010}), and have led to the hypothesis that there may be a `mass gap' $M_{\rm gap}\sim [2,5]M_\odot$ between the most massive neutron stars and the less massive black holes. However, we cannot entirely ignore the possibility that lower mass black holes exist, either in the mass gap or even at masses $M_{\rm BH} \lesssim 2M_\odot$. Accordingly, for GW170817, the idea that one of the two merging objects was a low-mass black hole instead of a neutron star cannot be discounted. More generally, the possible existence of low-mass black holes should be kept in mind when interpreting the larger number of expected joint GW-EM observations that will soon be at our disposal, whether the inferred mass of an object is below $2M_\odot$ or within the potential mass gap.

To understand what a low-mass neutron star-black hole (NSBH) merger would look like to GW and EM observers, we need general relativistic simulations of these systems. In particular, we need to understand the properties of the post-merger remnant and of any matter unbound during the merger, as such predictions are critical to model the EM signals powered by NSBH binaries. Low-mass NSBH merger simulations have recently been performed to calibrate NSBH GW templates~\cite{Hinderer:2016eia}. However, these simulations use equations of state that are far too simple to reliably model the outcome of the merger. Simulations with realistic equations of state have quite reasonably focused on more massive black holes ($M_{\rm BH}\sim (4-10)M_\odot$) (see e.g.~\cite{Etienne:2007jg,Duez:2008rb,Kyutoku:2010zd,Chawla:2010sw,Paschalidis2014,kyutoku:2015,Kiuchi:2015qua,Kawaguchi:2015,FoucartBhNs2016}), while our understanding of near equal-mass NSBH mergers has so far come from extrapolation of these simulation results to the equal mass regime. 

In this manuscript, we perform simulations of NSBH mergers at mass ratio $Q=(1-1.89)$, using a composition and temperature dependent equation of state (DD2~\cite{Hempel:2011mk}). We consider mass ratios $Q=1,1.2$ that most easily mimic galactic NSNS systems, as well as two higher mass ratio systems chosen for their consistency with GW170817. We show that while some extrapolations of existing fitting formulae to low mass ratio work quite well (for the ejected mass, and final black hole mass), others lead to inaccurate results (remnant disk mass, black hole spin, ejecta velocity). Our simulations provide an important point of reference to calibrate improved analytical formulae allowing us to model NSBH binaries at low mass ratio, as illustrated by our recent update to analytical predictions for the amount of mass remaining outside of the black hole after merger~\cite{Foucart:2018rjc}, and in our ability to study whether a given merger is a NSNS, NSBH, or binary black hole merger (see e.g.~\cite{2018arXiv180803836H,2019arXiv190106052C} for studies of GW170817 as a potential NSBH merger using these analytical predictions).

\section{Methods}

\subsection{Initial data}

\begin{table}
\begin{center}
\caption{Initial parameters of the binaries studied in this paper. $M_{\rm BH}$ is the Christodoulou mass of the black hole,
$M_{\rm NS}$ the ADM mass of an isolated neutron star with the same equation of state and baryon mass as the neutron star under
consideration, $\chi_{\rm BH}$ is the dimensionless spin of the black hole, $\Omega_0$ is the initial angular velocity, and $M=M_{\rm BH}+M_{\rm NS}$. $\Delta x_{\rm dis}$ is the typical grid resolution for the finest level of refinement used during the disruption of the neutron star (see Sec.~\ref{sec:grid} for more detail on the grid structure).}
{
\begin{tabular}{c||c|c|c|c|c}
Model & $M_{\rm BH}\,(M_\odot)$ & $M_{\rm NS}\,(M_\odot)$ & $\chi_{\rm BH}$ & $\Omega_0 M$ & $\Delta x_{\rm dis}$ (m) \\
\hline
B144N144 & 1.44 & 1.44 & 0.00  & 0.0233 & 235\\
\hline
B144N120-lr & 1.44 & 1.20 & 0.00  & 0.0206 & 295\\
B144N120 & 1.44 & 1.20 & 0.00  & 0.0206 & 235\\
\hline
B160N116 & 1.16 & 1.60 & 0.00  & 0.0218 & 235\\
\hline
B189N100 & 1.00 & 1.89 & 0.15  & 0.0232 & 235\\
\end{tabular}
\label{tab:ID}
}
\end{center}
\end{table}

We prepare initial data using our in-house Spells solver~\cite{Pfeiffer2003,FoucartEtAl:2008}. We first obtain initial data for NSBH binaries in quasi-circular orbit, then perform one iteration of the eccentricity reduction algorithm developed by Pfeiffer et al.~\cite{Pfeiffer-Brown-etal:2007} to obtain systems with residual eccentricity $e\sim 10^{-3}$. The neutron stars are initially in hydrostatic equilibrium, and have negligible spin. We consider 4 different configurations, listed in Table~\ref{tab:ID}. Two simulations are meant to mimic `average' NSNS binaries: an equal mass, non spining system ($M_{\rm BH}=M_{\rm NS}=1.44M_\odot$, with $M_{\rm BH}$ the Christodoulou mass of the black hole and $M_{\rm NS}$ the ADM mass of an isolated neutron star with the same baryon mass as the neutron star evolved in our simulation), and a slightly asymmetric system with $M_{\rm NS}=1.2M_\odot$, $M_{\rm BH}=1.44M_\odot$. These masses are within the most common range of observed masses for neutron stars in our galaxy. The main objective of these simulations is to understand the dynamics of near-equal mass NSBH binaries, and to allow us to extend to low mass ratios existing fitting formulae developed for the post-merger remant~\cite{Foucart2012,Pannarale:2014}, and for the amount of dynamical ejecta unbound by a merger~\cite{Kawaguchi:2016}. 

Two additional simulations are chosen to study `extreme' configurations compatible with GW170817. We consider that binary parameters are compatible with GW170817 if the chirp mass, mass ratio, effective spin, and effective tidal deformability of the binary lie within the 90\% confidence region published by the LVC~\cite{GW170817-PE}. The LVC performed parameter estimation using two different priors: negligible spins, or arbitrary spins (hereafter `low spin prior' and `high spin prior'). We simulate the most asymmetric mass ratio compatible with the low-spin prior, and the most asymmetric mass ratio compatible with the high-spin prior. Taking the more massive object to be a black hole, and assuming that the effective spin parameter measured through GWs,
\beq
\chi_{\rm eff} = \frac{M_{\rm NS} \chi_{z,\rm NS} + M_{\rm BH} \chi_{z,\rm BH}}{M_{\rm NS}+M_{\rm BH}}
\eeq
(with $\chi_{z,{\rm BH/NS}}$ the aligned component of the spin of the compact objects), is entirely due to the black hole spin, i.e. that $\chi_{z,\rm NS}=0$, we get $M_{\rm BH}=1.6M_\odot$, $M_{\rm NS}=1.16M_\odot$ in the first case (no spins), and $M_{\rm BH}=1.89M_\odot$, $M_{\rm NS}=1M_\odot$ for the second case (with a black hole dimensionless spin $\chi_{\rm BH}=0.15$, aligned with the orbital angular momentum). 

In the rest of this text, we label the simulations through the masses of the two compact objects, i.e. B160N116 corresponds to $M_{\rm BH}=1.6M_\odot$, $M_{\rm NS}=1.16M_\odot$. The longest simulation (B144N120) is initialized $\sim 8$ orbits before merger, and the shortest (B144N144) $\sim 6.5$ orbits before merger. All simulations use the DD2 equation of state~\cite{Hempel:2011mk}, a temperature and composition dependent equation of state that remains close to known nuclear physics constraints and is compatible with the existence of a $2M_\odot$ neutron star (the maximum mass of a neutron star with the DD2 equation of state is $\sim 2.42M_\odot$). The DD2 equation of state lies at the stiffer end of what is allowed by GW170817 if that event is a NSNS binary, but comfortably within the allowed range of tidal deformability if GW170817 is a NSBH binary. The dimensionless tidal deformability $\tilde \Lambda$ (the parameter actually measured through gravitational waves~\cite{GW170817-PE}) is $\tilde \Lambda\sim (605,550,600,295)$ for the four systems considered here (going from lowest to highest neutron star mass). Observational bounds require $\tilde \Lambda \lesssim 800$.  A summary of the properties of the neutron stars evolved in this paper is provided in Table~\ref{tab:NS}.

A brief description of cases B144N120 and B144N144 was already provided in~\cite{2018arXiv180803836H}, and these simulations were used to calibrate our most recent fitting formula for the mass of the matter remaining outside of the black hole after a BHNS merger~\cite{Foucart:2018rjc}. This manuscript presents a more complete description of these two simulations. The other two cases are reported here for the first time.

\begin{table}
\begin{center}
\caption{Properties of the simulated neutron stars. All neutron stars are modeled using the DD2 equation of state. $\Lambda=(2/3) k_2 (R_{\rm NS}c^2)^5/(GM_{\rm NS})^5$ is the tidal deformability and $k_2$ the Love number.}
{
\begin{tabular}{|c||c|c|c|c|}
\hline
NS Mass [$M_\odot$] & 1 & 1.16 & 1.2  & 1.44\\
\hline
Radius [km] & 13.1 &  13.1 & 13.1  & 13.2 \\
$\Lambda$ & 4190 & 1950 &  1630 &  590\\
\hline
\end{tabular}
\label{tab:NS}
}
\end{center}
\end{table}

\subsection{Evolution algorithm}

We evolve these NSBH binaries using the SpEC code~\cite{SpECwebsite}, following their evolution through late inspiral, merger, and the first $10\,{\rm ms}$ of post-merger evolution. SpEC evolves the equations of general relativity in the generalized harmonic formulation~\cite{Lindblom:2007} on a pseudo-spectral grid, and the general relativistic equations of hydrodynamics using shock-capturing finite volume methods~\cite{Duez:2008rb,Foucart:2013a}. In this work, we use the WENO5 algorithm to reconstruct fluid variables from cell centers to cell faces~\cite{Liu1994200,Jiang1996202,Borges}, and HLL fluxes as approximate solutions to the Riemann problem at faces~\cite{HLL}. 

We also evolve neutrinos with an approximate, gray two-moment transport scheme~\cite{1981MNRAS.194..439T,shibata:11,Cardall2013}. In the two-moment scheme, the energy and momentum density of each species of neutrinos are evolved on the grid. We then use the Minerbo analytical closure to provide the pressure tensor of the neutrinos~\cite{Minerbo1978}. The implementation of the two-moment transport into SpEC is described in~\cite{FoucartM1:2015,Foucart:2016rxm}. Recent studies using more advanced neutrino transport methods indicate that our two-moment scheme should be reasonably accurate except for its inability to properly capture energy deposition from neutrino-antineutrino pair annihilation in the polar regions -- a process that is entirely neglected in our simulations. We consider 3 distinct neutrino species: electron neutrinos $\nu_e$, electron antineutrinos $\bar \nu_e$, and a heavy-lepton neutrino species that regroups all other types of neutrinos $\nu_x = (\nu_\mu,\bar\nu_\mu,\nu_\tau,\bar\nu_\tau)$. 

In NSBH mergers, the main role of neutrinos is to cool the remnant accretion disk and to modify its composition. Neutrinos also play a subdominant role in driving post-merger disk winds, and a critical role in setting the composition of any disk wind. The dominant drivers of outflows are hydrodynamical processes during merger (tidal disruption, circularization) and magnetic processes after merger. While our simulations capture the first type of outflows, we do not evolve magnetic fields. Accordingly, we stop all simulations $\sim 10\,{\rm ms}$ after merger. Over longer timescales, magnetic fields are necessary to properly capture the evolution of the post-merger remnant and to assess the potential of a given post-merger remnant to power a relativistic jet and a gamma-ray burst (see e.g.~\cite{Nouri:2017fvh} for comparison of the evolution of remnant disks with and without magnetic fields, ~\cite{Chawla:2010sw,2012PhRvD..85f4029E,Paschalidis2014,Kiuchi:2015qua} for merger simulations including magnetic fields, and~\cite{Siegel:2017nub,Fernandez:2018kax} for longer evolutions of the post-merger remnant).

\subsection{Numerical grids}
\label{sec:grid}

For all simulations, the spectral grid used to evolve Einstein's equations is composed of a ball at the center of the neutron star, spherical shells around that ball and around the black hole, spherical shells far away from the compact objects ($>2.5$ orbital separations), and finally distorted cylinders connecting these three regions. The number of basis functions used by the pseudo-spectral code on each element is chosen adaptively to maintain a fixed truncation error, as described in~\cite{Szilagyi:2014fna,Foucart:2013a}. The equations of hydrodynamics and two-moment neutrino transport are evolved on a cartesian grid, with initial grid spacing provided in Table~\ref{tab:ID}. As the compact objects are kept fixed on the numerical grid, but the binary separation actually decreases over time, the grid spacing decreases as the evolution progress. Whenever the grid spacing decreases by $20\%$, we interpolate onto a new evolution grid, restoring the original grid spacing. The spatial extents of the cartesian grid are chosen adaptively, and cover all regions with a significant amount of matter ($\rho>6 \times 10^{9}\,{\rm g/cm^3}$ within the initial orbit of the binary, with that threshold dropping as $\sim r^{-3}$ at larger distances). We monitor mass losses at the boundary of the computational domain, and find that this prescription limits them to less than $10^{-4}M_\odot$.

After disruption of the neutron star, we use nested grids. Each level of refinement has $252^3$ cells. The finest grid has the same resolution as during inspiral, and the grid spacing is multiplied by a factor of two between refinement levels. Once the densest point in the simulation is in the remnant disk rather than close to the horizon of the black hole, we remove the finest level of refinement, to save computational resources. Indeed, the most relevant scale for the post-merger remnant is the radius of the disk, which is larger than the size of the original neutron star. In the absence of magnetic fields, we do not need to resolve smaller physical scales after the formation of an accretion disk.

\subsection{Error estimates}

The grid resolution used for this study is comparable to recent NSBH simulations with SpEC~\cite{FoucartBhNs2016,Brege:2018kii}. In~\cite{FoucartBhNs2016}, convergence tests for higher mass ratio systems showed relative errors of $\sim 20\%$ in the measured mass of the dynamical ejecta, $\sim 10\%$ in the mass of the remnant accretion disk, and $<1\%$ in the properties of the black hole. High mass ratio systems are generally more demanding numerically, due to the formation of thin, hard-to-resolve accretion streams during tidal disruption~\cite{Foucart:2014nda,FoucartBhNs2016}. The simulations presented here should have at worse comparable errors.

To verify this, simulation B144-N120 was performed at a lower resolution up to the end of the simulation ($\Delta x=295\,{\rm m}$ instead of $\Delta x=235\,{\rm m}$), and at a finer resolution ($\Delta x=190\,{\rm m}$) up to the end of the disruption of the neutron star ($1\,{\rm ms}$ after merger). The 3 resolutions show better than second-order convergence, and the errors are consistent with~\cite{FoucartBhNs2016} (or slightly better), with the exception of the mass of dynamical ejecta. The dynamical ejecta produced in this simulation is too small to be resolved ($\sim 0.001M_\odot$). The highest mass ratio simulation performed here (B189N100) is the only configuration for which enough dynamical ejecta is produced to expect $\sim 20\%$ relative errors. The ejecta is only qualitatively captured in B160N116, and it is consistent with no ejecta in B144N120 and B144N144.

\section{Results}

\subsection{Merger and Remnant Properties}

\begin{figure*}
\begin{center}
\includegraphics[width=.99\textwidth]{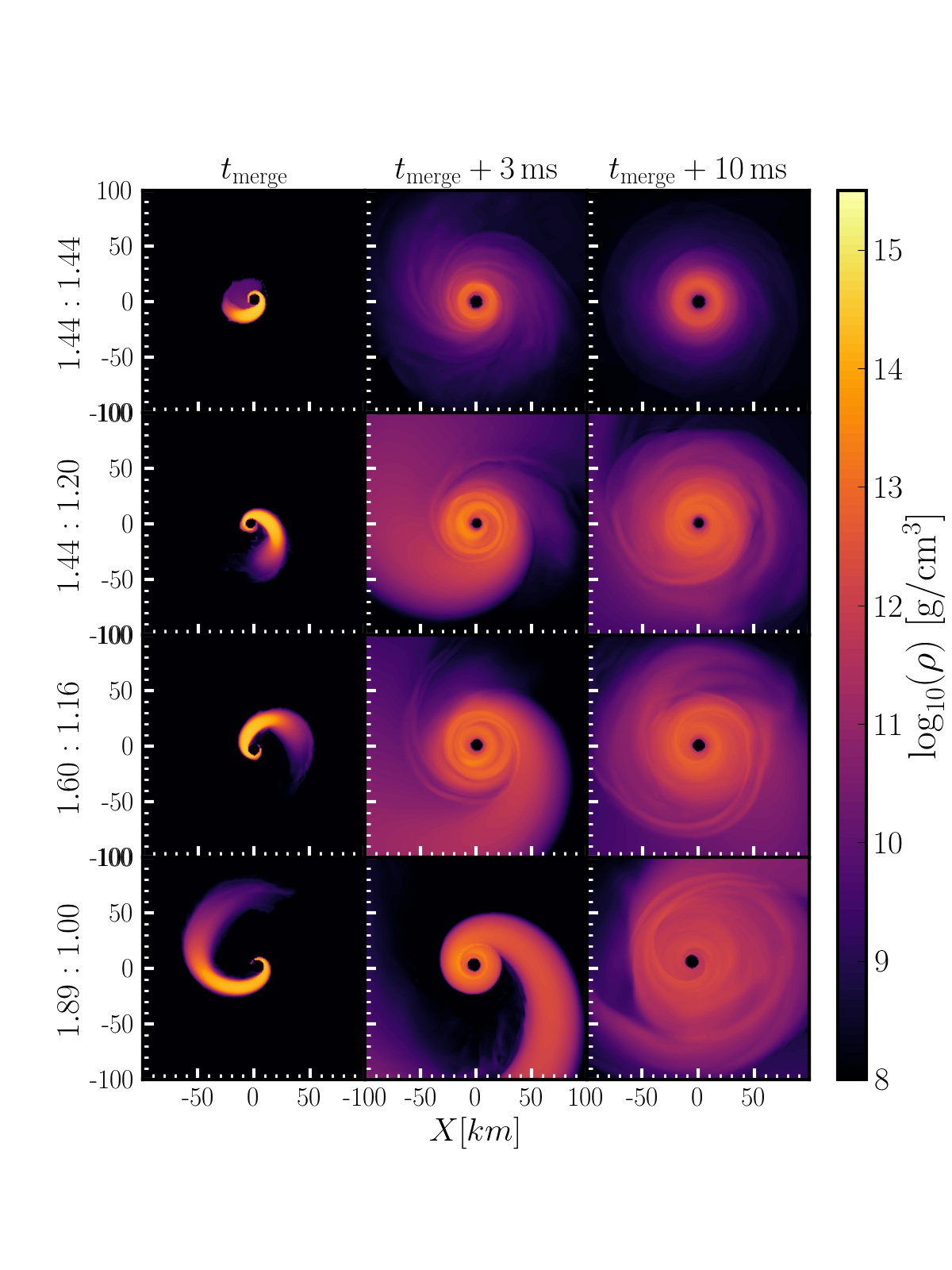}
\caption{Baryon density in the equatorial plane of our simulations. The left panel shows results at the time of merger (when $50\%$ of the mass of the neutron star has been
accreted by the black hole), the middle panel $3\,{\rm ms}$ later, and the right panel $10\,{\rm ms}$ later, at the end of our simulations. From top to bottom, we show all four configurations starting with the equal-mass system and moving towards the most asymmetric mass ratio.}
\label{fig:DensityVis}
\end{center}
\end{figure*}

\begin{table}
\begin{center}
\caption{Outcome of the simulations. $M_{\rm BH}^f$ and $\chi_{\rm BH}^f$ are the mass and dimensionless spin of the remnant black hole.
$M_{\rm ej}$ is the amount of mass ejected by the merger (matter with $h u_t<-1$). $M_{\rm rem}$ the baryon mass remaining outside of the black hole.
All quantities are measured $10\,{\rm ms}$ after merger.}
{
\begin{tabular}{c||c|c|c|c}
Model & $M_{\rm BH}^f\,(M_\odot)$ & $\chi_{\rm BH}^f$ & $M_{\rm ej}\,(M_\odot)$ & $M_{\rm rem}\,(M_\odot)$  \\
\hline
B144N144 & 2.81 & 0.84 & $ 0.0002$  &  0.03 \\
\hline
B144N120-lr & 2.49 & 0.80 & 0.002  & 0.13 \\
B144N120 & 2.49 & 0.80 & 0.001  & 0.12 \\
\hline
B160N116 &  2.59 & 0.76 & 0.004 & 0.13\\
\hline
B189N100 & 2.68 & 0.70 &  0.05 & 0.18 \\
\end{tabular}
\label{tab:results}
}
\end{center}
\end{table}

The overall dynamics of the four configurations studied here are visible in Fig.~\ref{fig:DensityVis}, while global properties of the post-merger remnant and dynamical ejecta are listed in Table~\ref{tab:results}. We observe three fairly different outcomes for these low-mass systems. 

In the equal mass configuration, B144N144, the neutron star is barely disrupted by the tidal potential of the black hole. Only a small amount of matter remains outside of the black hole after merger ($\sim 0.03M_\odot$). There is no matter unbound by the merger, and the small amount of mass remaining outside of the black hole is insufficient to explain the type of kilonova observed following GW170817. This is a surprising results: from preexisting fitting formulae~\cite{Foucart2012} we expected a remnant mass of $\sim 0.22M_\odot$, probably {\it too high} to be compatible with GW170817 (assuming $\sim 40\%$ of the disk being unbound during post-merger evolution~\cite{Siegel:2017nub,Fernandez:2018kax}). To account for this discrepancy, we have now developed a new fitting formula extending to near equal-mass NSBH binaries~\cite{Foucart:2018rjc}. This formula still somewhat overestimates the mass outside of the black hole for this configuration (prediction of $0.06M_\odot$), but by much less than the original result. Considering that the DD2 equation of state is already quite stiff, and thus other equations of state would lead to less massive remnant disks, our results indicate that an equal mass NSBH merger is strongly disfavored as the progenitor of GW170817.

The remnant disk rapidly circularizes, with hydrodynamics shocks increasing the temperature of the remnant to $\langle T\rangle \sim 4\,{\rm MeV}$ within $3\,{\rm ms}$ of the merger
\footnote{Here and in the rest of the text, $\langle X \rangle$ denotes the density-weighted average of the variable $X$}. This leads to rapid protonization of the remnant disk: the equilibrium $Y_e$ of the remnant is higher than the $Y_e$ of the neutron star, and neutrino emissions thus drive $Y_e$ up, to $\langle Y_e \rangle\sim 0.15$ ($4\,{\rm ms}$ after merger). After that circularization phase, energy losses to neutrinos cause the disk to become more compact (but not cooler), and its composition to become slightly more neutron rich. By the end of the simulation, most of the material is in a compact torus with peak density at $\sim (20-30)\,{\rm km}$, $\langle Y_e \rangle \sim 0.12$, and $\langle T\rangle\sim 4\,{\rm MeV}$. This evolution is very similar to post-merger evolutions at higher mass ratios~\cite{Deaton2013,Foucart:2014nda}, albeit the evolution of the disk happens on a shorter time scale in this simulation.

The two median cases, B144N120 and B160N116, are strikingly similar despite have different neutron star masses, neutron star compactions ($C_{\rm NS}=M_{\rm NS}/R_{\rm NS}$), and black hole masses. Both lead to strong disruption of the neutron star and leave $\sim (0.12-0.13) M_\odot$ of material outside of the black hole $10\,{\rm ms}$ after merger, without much mass ejection. The similarity between these two cases is predicted by both the old and new fitting formulae for the remnant mass, but the old results again overestimated the matter left outside of the black hole (predicted $\sim [0.23-0.24]M_\odot$), while the new fitting formula is extremely accurate (predicted $\sim [0.13-0.15]M_\odot$). This is of particular interest in the context of GW170817, as a post-merger disk of mass $M_{\rm disk} \sim 0.1M_\odot$ is probably what is needed to eject the right amount of mass to power the observed kilonova. There are other difficulties that may arise when trying to explain GW170817 as a NSBH merger, but the mass budget of the outflows at least is consistent with GW170817 (see~\cite{2018arXiv180803836H} for a more in-depth discussion of GW170817 as a NSBH merger).

For these two cases, the remnant disk is not as rapidly circularized as in the first simulation. By the end of the simulation, the temperature of the disk is still increasing, to $\langle T\rangle = 3.3\,{\rm MeV}$ (resp. $\langle T\rangle = 2.6\,{\rm MeV}$) for B144N120 (resp B160N116). As a result of the lower temperature, the disk remains very neutron rich ($\langle Y_e \rangle \lesssim 0.1$), although this would probably change over longer time scales or in the presence of magnetically-driven turbulent heating. The final remnant is still far from axisymmetric, but it is compact: most of the matter is within $\sim 50\,{\rm km}$ of the black hole.

Finally, B189N100, the more extreme mass ratio with a slowly spinning black hole, shows strong disruption of the neutron star, the ejection of a significant amount of material in an unbound tidal tail ($\sim [0.03-0.05]M_\odot$, see next section), and again $\sim (0.13-0.15)M_\odot$ of bound material at the end of the simulation. This simulation is also a success for our new fitting formula~\cite{Foucart:2018rjc}: it overestimates the mass remaining outside of the black hole by only $10\%$ (vs $30\%$ for the old formula). In the context of GW170817, case B189N100 is disfavored, as too much mass is ejected through a combination of dynamical ejecta and later disk outflows --- but a more compact neutron star may be an acceptable alternative for similar binary parameters. The matter remaining outside of the black hole remains quite cold ($\langle T \rangle =1.6\,{\rm MeV}$) and its $Y_e$ does not significantly increase over the duration of the simulation ($\langle Y_e\rangle= 0.06$ at the end of the simulation).

The properties of the final black hole are also important to characterize the post-merger remnant. We compare our numerical results with the analytical predictions of~\cite{Pannarale:2014} for the final mass and spin of the black hole. We could use directly the results of~\cite{Pannarale:2014}, but that manuscript made use of our old fitting formula for the baryon mass outside of the black hole after merger~\cite{Foucart2012}, which is unreliable in this regime. As~\cite{Foucart2012} overestimates the torus mass,~\cite{Pannarale:2014} naturally underestimates the remnant black hole mass. Using the updated formula~\cite{Foucart:2018rjc} instead, the remnant black hole mass is reasonably well predicted - with an error of $0.03M_\odot$ for the equal mass case, and of $\lesssim 0.01M_\odot$ for all other cases. The spin of the black hole, on the other hand, is more problematic. Going from the system with highest neutron star mass to the system with lowest neutron star mass, the black hole spins predicted by~\cite{Pannarale:2014} are $\chi_f=0.72,0.70,0.68,0.68$, while numerical results are $\chi_f=0.84,0.80,0.76,0.70$, i.e. the highest mass ratio system is the only one reasonably well modeled by the analytical formula. This may be because the analytical formula adds to the black hole spin the angular momentum of the accreting matter at the innermost stable orbit of the {\it final} black hole, while some of the matter presumably plunged from the innermost stable orbit of the {\it initial} black hole -- and the difference between these two assumptions is quite large for near equal mass systems. An updated analytical formula for the final black hole spin is thus necessary for reliable predictions in the near equal-mass regime.

\subsection{Matter outflows}

Another important output of our simulations is the amount of matter unbound through tidal disruption of the neutron star. Indeed, that unbound material can play a significant role in the production of a kilonova days to weeks after the GW signal. For neutron stars merging with typical stellar mass black holes ($M_{\rm BH}>5M_\odot$), we know that the neutron star either plunges directly into the black hole, producing neither ejecta nor disk, or is disrupted and ejects large amounts of neutron-rich material (typically a few percents of a solar mass). We have already seen that the situation is quite different for our near equal-mass systems: neutron stars that clearly undergo tidal disruption end up producing a negligible amount of dynamical ejecta. 

To judge how uncommon that situation is, we consider the correlation between disk mass and ejecta mass found by Kyutoku et al.~\cite{kyutoku:2015}: binaries with remnant disk mass of $\sim 0.1M_\odot$ typically produce $\sim 0.01M_\odot$ of dynamical ejecta. There is however a significant scatter in that relation for disk masses $\lesssim 0.1M_\odot$: for a small number of binaries with mass ratios $Q\sim 3$, Kyutoku et al.~\cite{kyutoku:2015} find ejected masses only slightly higher than those found in our near equal-mass simulations, with disk masses of $\sim 0.1M_\odot$. 
We can also look at the fitting formula developed for the amount of ejected mass by Kawaguchi et al.~\cite{Kawaguchi:2016}. As for our outdated fitting formula for the mass remaining outside of the black hole~\cite{Foucart2012}, the formula from~\cite{Kawaguchi:2016} was calibrated to simulations at mass ratios $Q=3-7$, and has no particular reason to remain valid at lower mass ratios. However, it does end up working remarkably well for $Q\sim 1$. The formula correctly predicts the lack of ejecta for the equal mass system, and predicts ejected masses of $\sim (0.012-0.013)M_\odot$ for the two intermediate systems B144N120 and B160N116. As the fitting formula is accurate to $\sim 0.01M_\odot$, this is consistent with our numerical results. The formula also correctly captures the rapid rise in the ejected mass for the most asymmetric system, predicting an ejected mass of $0.06M_\odot$, close to the numerical result of $0.05M_\odot$ (if we use the same criteria to compute the unbound material, see below). The success of this fitting formula at low mass ratios (and the lack of accuracy of~\cite{Foucart2012} in that same regime) may be due to the use of a more complex dependence of the fitting formula in the mass ratio in the ejecta model.
Overall, we thus see that while this combination of negligible ejecta and massive remnant disk is not common for higher mass ratio binaries, it is neither unprecedented nor particularly unexpected given the predictions of Kawaguchi et al.~\cite{Kawaguchi:2016}.

\begin{table}
\begin{center}
\caption{Ejecta properties for the 2 simulations producing a measurable outflow mass. $M_{\rm ej,hu_t}$ is the ejected mass using the $hu_t<-1$ criteria,
$M_{\rm ej,u_t}$ is the ejected mass using the $u_t<-1$ criteria, $\langle Y_e \rangle$ is the average electron fraction of the ejecta, $\langle v \rangle$ its average velocity, and $T_{\rm kin}$ its total kinetic energy (all computed for the $hu_t<-1$ criteria).}
{
\begin{tabular}{c||c|c|c|c|c|}
Model & $M_{\rm ej,hu_t}$ & $M_{\rm ej,u_t}$ & $\langle Y_e \rangle$ & $\langle v \rangle$ & $T_{\rm kin}$ (ergs)  \\
\hline
B160N116 &  $0.004M_\odot$ & $0.001M_\odot$ & 0.05 & 0.10c & 4.6e49\\
\hline
B189N100 & $0.05M_\odot$ & $0.03M_\odot$ &  0.05 & 0.14c & 1.0e51\\
\end{tabular}
\label{tab:ej}
}
\end{center}
\end{table}

For our higher mass ratio simulations, B160N116 and B189N100, we resolve the dynamical ejecta and can look at its properties in more detail (see Table~\ref{tab:ej}). First, we note that in the table and in our discussions so far, we have computed the ejected mass using the `Bernoulli' criteria $h u_t<-1$, with $h$ the specific enthalpy of the fluid and $u_t$ the time component of the 4-velocity one-form. This typically overestimates the amount of unbound ejecta, because it assumes that all thermal energy and all energy released through r-process nucleosynthesis is transformed into kinetic energy. An alternative method is to require $u_t<-1$, which assumes that none of the thermal and r-process energy is transformed into kinetic energy. 

\begin{figure}
\begin{center}
\includegraphics[width=.99\columnwidth]{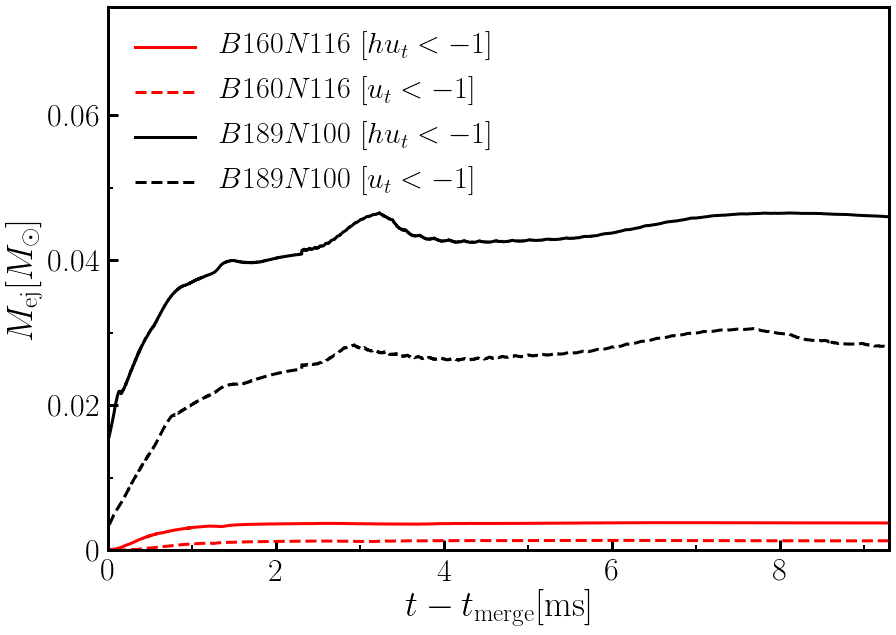}
\caption{Total mass flagged as unbound as a function of time for the two configurations producing a measurable amount of dynamical ejecta. We show results for the two criteria $hu_t<-1$ and $u_t<-1$, which converge to different answers because $\lim_{\rho\rightarrow 0} h \neq 1$ for the DD2 equation of state (see text). After $\sim 3{\rm ms}$, the error made by measuring the mass of dynamical ejecta at finite radius is typically smaller than numerical errors, and than the errors caused by assuming NSE in the fluid.}
\label{fig:Outflow}
\end{center}
\end{figure}

For the cold ejecta produced in black hole-neutron star mergers, the thermalization of the energy released through r-process nucleosynthesis is by far the most important of these two factors, and the only correct way to treat it would be to use an equation of state that does not assume nuclear statistical equilibrium (NSE) and follows the number density of each type of atomic nuclei. This is beyond what our code can currently do. The thermalization of the energy released through r-process nucleosynthesis has been studied in details on timescales relevant for kilonova observations (days)~\cite{Barnes:2016}, but not on the $\sim$second time scales where most of the energy is released. A reasonable estimate is that about $50\%$ of that energy is released in neutrinos and escapes, while the other $50\%$ thermalizes~\cite{2010MNRAS.402.2771M}. In that case, the correct answer for the ejected mass would lie about half-way between the predictions of the $hu_t<-1$ and $u_t<-1$ criteria. The $u_t$ ($hu_t$) criteria predicts $0.03M_\odot$ ($0.05M_\odot$) of ejected mass for B189N100 and $0.001M_\odot$ ($0.004M_\odot$) of ejected mass for B160N116. We thus see that out-of-NSE evolution is a source of error at least comparable to current numerical errors in NSBH simulations. The error due to out-of-NSE evolution and the numerical error are here more important than the uncertainty due to the measurement of the ejected mass at a finite radius : Fig.~\ref{fig:Outflow} shows that the amount of unbound mass measured using either criteria does not change for the last few milliseconds of our simulation, as the unbound material moves away from the remnant.

In both simulations, nearly all of the ejecta is extremely neutron rich ($Y_e<0.1$) and, at the end of the simulation, cold ($T\lesssim 0.1\,{\rm MeV}$). This is typical for the dynamical ejecta from NSBH binaries, and will inevitably lead to the production of large amounts of lanthanides and actinides during r-process nucleosynthesis. The velocity of the ejecta is more surprising, with the average velocity being $\langle v \rangle=0.1c$ for B160N116 and $\langle v \rangle=0.14c$ for B189N100. This is significantly slower than in NSBH simulations performed at higher mass ratios. For mass ratios $Q\sim 3-7$, we typically have $\langle v \rangle\sim (0.2-0.3)c$~\cite{Kawaguchi:2016}. We can also extrapolate to $Q\sim 1-2$ the fitting formula of Kawaguchi et al.~\cite{Kawaguchi:2016} for the velocity, and find predicted velocities of $\langle v \rangle\sim (0.20-0.22)c$. We thus find that while predictions for the ejected mass extrapolate well to the equal mass regime, predictions for the velocity of the ejecta do not. This has important consequences for the observational properties of the dynamical ejecta from low-mass NSBH binaries: a neutron-rich, low-velocity ejecta is often associated with disk outflows, but we see here that it can in fact be produced by a near equal-mass NSBH systems. 

Over longer time scales ($\sim [0.01-1]\,{\rm s}$), the main source of outflows in the post-merger remnant is nearly certainly going to be magnetically driven~\cite{Siegel:2017nub,Fernandez:2018kax}. These magnetically driven winds, which cannot be captured by our simulations but most likely have a total mass of $\sim (20-50)\%$ of the post-merger disk mass, should certainly be taken into account when modeling kilonovae associated with low-mass BHNS mergers. Even in the absence of magnetic fields, however, neutrino absorption in low-density regions above the disk can lead to the production of a neutrino-driven wind~\cite{Dessart2009,Perego2014,Just2014}.
We confirm that, by the end of our simulations, such a wind is present. The outflow rate is quite low, $(0.01-0.04)M_\odot/s$, , and its contribution to the total mass budget of the outflows is thus small. The neutrino driven wind is neutron poor ($Y_e>0.35$), except for case B189N100. In that case, the post-merger disk is colder, neutrino irradiation of the wind is not as significant, and the electron fraction of the late-time outflows is still $Y_e\lesssim 0.3$. We note however that the composition of the neutrino-driven outflows observed in our simulations may not be representative of the composition of post-merger outflows, as magnetically driven winds are likely to be denser and faster than the outflows observed in our simulations. Finally, we also emphasize that the post-merger remnants produce in our simulations may power relativistic jets and short gamma-ray bursts, but that the lack of magnetic field in our simulations makes it impossible to study jet production here.

\section{Discussion}

In this manuscript, we perform the first general relativistic simulations of near equal-mass, quasi-circular NSBH binaries going beyond the use of unrealistic ideal gas equations of state. Current observations in the Milky Way~\cite{Ozel:2010} and through gravitational waves~\cite{LIGOScientific:2018jsj} favor higher mass black holes, but the existence of solar mass black holes, or alternatively of black holes in the `mass gap' $[2,5]M_\odot$, cannot entirely be ruled out. A solar mass black hole in a binary system with a neutron star companion could mimic the observable properties of a NSNS binary, affecting our interpretation of current and upcoming observations of compact binary mergers. To properly understand current and upcoming observations of binary mergers involving neutron stars, we should thus carefully model the observable properties of low-mass NSBH binaries.

Our simulations show that these systems produce post-merger remnant disks that are significantly less massive than previously expected, a conclusion that has already led us to update analytical predictions for the outcome of NSBH mergers~\cite{Foucart:2018rjc}. In the context of GW170817, our results also show that NSBH mergers can reproduce both the observed gravitational wave signal and the inferred mass budget of the outflows produced by that merger. In separate work, we showed that our simulation results imply that large neutron stars are favored in the (arguably unlikely) event that GW170817 is a NSBH merger~\cite{2018arXiv180803836H}, in contrast with results derived assuming a NSNS merger. Our updated model for the outcome of NSBH mergers, as well as results presented here for the final mass and spin of the remnant black hole and the properties of the dynamical ejecta produced in low-mass NSBH mergers, can also play an important role in the interpretation of the many NSNS/NSBH mergers expected during the upcoming O3 run of Advanced LIGO/Virgo. 

For the dynamical ejecta, we find that binaries with mass ratio $Q \lesssim 1.3$ produce nearly no dynamical ejecta, even for the relatively stiff equation of state considered here. At higher mass ratios, the observed ejected mass is consistent with predictions based on higher mass ratio simulations~\cite{Kawaguchi:2016}, but slower than predicted. In fact, we suggest that the low-velocity neutron rich dynamical ejecta produced in a low-mass NSBH merger may be difficult to distinguish from ejecta produced over the secular evolution of the remnant accretion disk, which may complicate the interpretation of future kilonova observations. 

Finally, we note that the remnant black hole itself has a remnant mass consistent with analytical predictions~\cite{Pannarale:2014}, but is spinning much faster than previously believed. The remnant black hole of our equal-mass NSBH merger has a dimensionless spin $\chi_{\rm BH}=0.84$, well above theoretical expectations for NSBH mergers or the spin of black holes resulting from NSNS mergers. 

\acknowledgments
The authors thank Tanja Hinderer as well as the members of the SXS collaboration for helpful discussions over the course of this project. 
F.F. gratefully acknowledges
support from NASA through grant 80NSSC18K0565,
and from the NSF through grant PHY-1806278
M.D. acknowledges support through
NSF Grant PHY-1806207.
S.M.N. is grateful for support from NWO
VIDI and TOP Grants of the Innovational Research Incentives
Scheme (Vernieuwingsimpuls) financed by the Netherlands
Organization for Scientific Research (NWO). 
H.P. gratefully acknowledges support from the NSERC Canada.
L.K. acknowledges support from NSF
grant PHY-1606654. M.S.
acknowledges support from NSF Grants PHY-170212 and
PHY-1708213.
L.K. and M.S. also thank the Sherman Fairchild Foundation for their support.
Computations were performed on the supercomputer Briar\'ee from the Universit\'e de Montr\'eal,
managed by Calcul Qu\'ebec and Compute Canada. The operation of these supercomputers is funded
by the Canada Foundation for Innovation (CFI), NanoQu\'ebec, RMGA and the Fonds de recherche du Qu\'ebec - Nature et
Technologie (FRQ-NT). Computations were also performed on the Zwicky and Wheeler clusters at Caltech, supported by the Sherman
Fairchild Foundation and by NSF award PHY-0960291.

\bibliographystyle{iopart-num}
\bibliography{References/References.bib}

\end{document}